# THE STRUCTURE OF COLD DARK MATTER HALOS

JULIO F. NAVARRO
*Steward Observatory*
*University of Arizona*
*Tucson, AZ 85721*
*U.S.A*

**Abstract.** High resolution N-body simulations show that the density profiles of dark matter halos formed in the standard CDM cosmogony can be fit accurately by scaling a simple "universal" profile. Regardless of their mass, halos are nearly isothermal over a large range in radius, but significantly shallower than $r^{-2}$ near the center and steeper than $r^{-2}$ in the outer regions. The characteristic overdensity of a halo correlates strongly with halo mass in a manner consistent with the mass dependence of the epoch of halo formation. Matching the shape of the rotation curves of disk galaxies with this halo structure requires (i) disk mass-to-light ratios to increase systematically with luminosity, (ii) halo circular velocities to be systematically lower than the disk rotation speed, and (iii) that the masses of halos surrounding bright galaxies depend only weakly on galaxy luminosity. This offers an attractive explanation for the puzzling lack of correlation between luminosity and dynamics in observed samples of binary galaxies and of satellite companions of bright spiral galaxies, suggesting that the structure of dark matter halos surrounding bright spirals is similar to that of cold dark matter halos.

1. **Introduction**

Dark matter halos are often modelled as isothermal potential wells whose depth is usually identified with the velocity dispersion of stars in spheroidal galaxies or with the disk rotation speed in spirals. Since velocity dispersion increases with galaxy luminosity, this implies that more massive halos should surround brighter galaxies, a hypothesis that, however, does not seem to be supported by observations. Studies of the dynamics of binary galaxies and of spiral/satellite samples have revealed a distinct lack of correlation between the dynamics and the luminosity of the system (White *et al.* 1983, Zaritsky & White 1994). Motivated by this disagreement, we have decided to investigate the structure of dark matter halos in the standard biased ($b = 1.6$) CDM cosmogony using high-resolution N-body simulations. A total of 19 halos with masses ranging from that of a rich galaxy cluster ($\sim 10^{15} M_\odot$) to that of a dwarf galaxies ($10^{11} M_\odot$) were selected from large cosmological simulations and resimulated individually with higher resolution. (I assume $H_0 = 50$ km s$^{-1}$ Mpc$^{-1}$ for all physical quantities quoted in this paper.) The numerical parameters were carefully chosen so that the numerical resolution of each simulated halo was the same, enabling meaningful comparisons between systems of widely different mass.

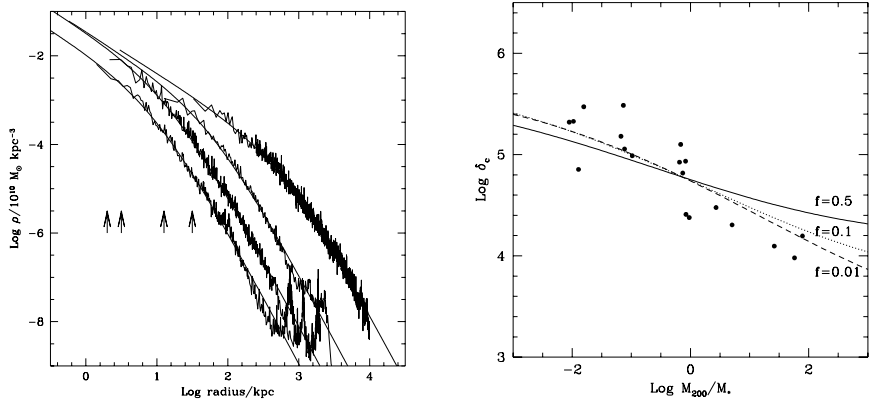

*Figure 1.* (**a**–left) The typical density profiles of CDM halos. The leftmost (rightmost) system has a mass $M_{200} = 3 \times 10^{11} M_\odot$ ($M_{200} = 3 \times 10^{15} M_\odot$). Arrows indicate the value of the gravitational softening in each simulation. (**b**–right) The characteristic overdensity of halos as a function of mass. Masses are expressed in units of the current non linear mass corresponding to the standard biased CDM spectrum, $M_* = 3.3 \times 10^{13} M_\odot$.

## 2. Results

Figure 1a shows the density profiles of four halos, spanning four orders of magnitude in mass. The solid curves are fits using a density profile of the form suggested by Navarro, Frenk & White (1995a),

$$\frac{\rho(r)}{\rho_{crit}} = \frac{\delta_c}{(r/r_s)(1 + r/r_s)^2}. \qquad (1)$$

Remarkably, all the profiles are very well fit by this simple functional form. (Here $\rho_{crit}$ is the critical density and $r_s$ is a "scale" radius.) If radii are expressed in units of the virial radius, $r_{200}$, (which determines the total mass of the system, $M_{200}$) there is a single free parameter in eq.(1); the characteristic overdensity $\delta_c$. Figure 1b shows that this overdensity correlates strongly with the mass of the halo; less massive systems tend to be denser than their larger counterparts. Such a trend is expected in hierarchical clustering scenarios such as CDM, since lower mass scales collapse at higher redshift and should therefore have typically higher characteristic densities. If we define the formation redshift of a halo of mass $M$ as the first time when half of its final mass was in progenitors with individual masses exceeding some fraction $f$ of $M$, we can compute analytically the expected dependence of $\delta_c$ on $M$ by assuming that characteristic overdensity of a halo merely reflects the density of the universe at the time of formation, i.e. $\delta_c(M) \propto (1 + z_{form}(M))^3$. The curves in Figure 1b show that such identification provides a good description of the results of the numerical experiments for various values of the parameter $f$, lending support to the conclusion that the characteristic density of a halo is determined primarily by its formation redshift. (See Navarro, Frenk & White 1995b for details.)

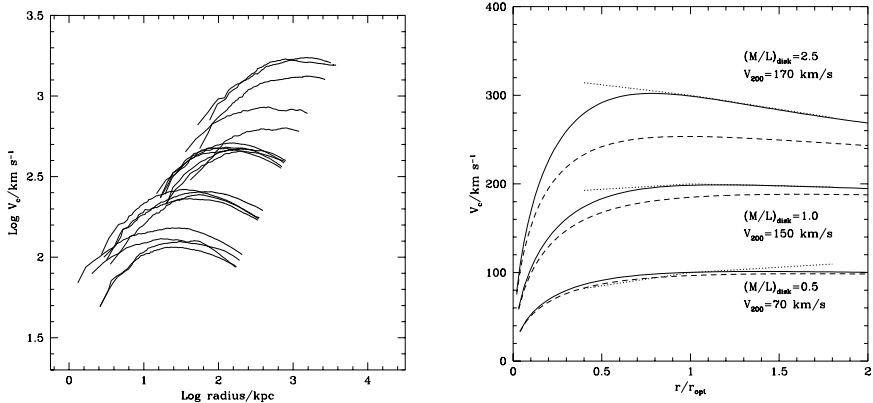

*Figure 2.* (**a**–left) The circular velocity as a function of radius for the 19 halos simulated in our series. The curves are truncated at the virial radius, $r_{200}$, of each system. (**b**–right) The result of matching the shape of observed disk rotation curves (dotted lines, from Persic & Salucci 1995). The dashed lines are the halo contribution to the total circular velocity (solid lines) in each case. Radii are expressed in units of the optical radius of each galaxy, defined as 3.2 times the exponential radial scalelength. $(M/L)_{disk}$ values are in the $B$-band.

The circular velocity, $V_c(r) = (GM(r)/r)^{1/2}$, of all 19 halos is shown in Figure 2a as a function of radius. Consistent with eq.(1), $V_c$ rises near the center and declines near the virial radius. Over a wide range in radius $V_c$ is almost constant, in agreement with the results of previous, lower resolution studies (Frenk *et al.* 1985, 1988, Quinn *et al.* 1986). Is the ra-

4                           JULIO F. NAVARRO

dial behaviour of $V_c$ consistent with the observed rotation curves of galaxy disks? To compute the rotation curve of disk galaxies with a given rotation speed forming within a CDM halo, only two parameters need to be specified; the disk stellar mass-to-light ratio, $(M/L)_{disk}$, and the halo mass or, equivalently, its circular velocity at the virial radius, $V_{200}^2 = GM_{200}/r_{200}$. (This is because, for a given rotation speed, observations constrain the luminosity and optical radius of a spiral disk, as well as the typical shape of the rotation curve.) Slight modifications to the halo structure caused by the presence of the disk can be taken into account by assuming that the halo responds adiabatically to the assembly of the galaxy (Barnes & White 1984, Blumenthal *et al.* 1986).

Figure 2b shows the values of these two parameters needed to match the typical rotation curves of spiral disks, as given by Persic & Salucci (1995). Some trends are clear; faster rotating (brighter) disks require larger disk mass-to-light ratios, $(M/L)_{disk} \propto L^{1/2}$, and the asymptotic halo circular velocity is systematically lower than that of the disk. Furthermore, bright galaxies, i.e. those with $V_{rot} > 200$ km/s, are surrounded by halos of very similar mass. Indeed, disks with rotation speeds 200 or 300 km/s should be surrounded by halos whose mean circular velocities differ by only $\sim 10\%$ and which can be up to a factor of two lower than the disk rotation speed. This result agrees well with the estimates of Zaritsky & White (1994), who found from their study of the dynamics of satellite galaxies that the average circular velocity of the halos of bright spirals is between 180 and 200 km/s. The weak correlation between halo and disk circular velocity shown in Fig. 2b also provides a simple explanation for the lack of correlation found between luminosity and dynamics of binary galaxies and satellite/primary pairs. This result is especially encouraging, since the halo properties we infer were chosen to match the shape of the inner rotation curves, and do not use any information about dynamics at the much larger radii probed by binaries or satellite companions. We conclude that the structure of dark halos surrounding spiral galaxies is quite similar to that of Cold Dark Matter halos.